\documentclass[letter,
               biblatex,     % biblatex is used
               keeplastbox,   % flushend option: not to un-indent last line in References
               ]{jacow}
\usepackage{pdfpages,multirow,ragged2e} %
\usepackage{svg}
\usepackage{listings}
\makeatletter%
	\ifboolexpr{bool{xetex}}
	 {\renewcommand{\Gin@extensions}{.pdf,%
	                    .png,.jpg,.bmp,.pict,.tif,.psd,.mac,.sga,.tga,.gif,%
	                    .eps,.ps,%
	                    }}{}
\makeatother
\ifboolexpr{bool{xetex} or bool{luatex}} % test for XeTeX/LuaTeX
 {}                                      % input encoding is utf8 by default
 {\usepackage[utf8]{inputenc}}           % switch to utf8

\usepackage[USenglish]{babel}
\usepackage{subcaption}

\ifboolexpr{bool{jacowbiblatex}}%
 {%
  \addbibresource{LLRF97.bib}
  \addbibresource{biblatex-examples.bib}
 }{}
\begin{document}

\title{Variational Autoencoeers for Noise Reduction in Industrial LLRF Systems}

\author{J. P.  Edelen, M. J. Henderson, J. Einstein-Curtis and C. C. Hall, RadiaSoft LLC, Boulder CO, USA, \\ J. A. Diaz Cruz and A. L. Edelen, SLAC, Menlo Park CA, USA}

\maketitle

\begin{abstract}
Industrial particle accelerators inherently operate in much dirtier environments than typical research accelerators. This leads to an increase in noise both in the RF system and in other electronic systems. Combined with the fact that industrial accelerators are mass produced, there is less attention given to optimizing the performance of an individual system. As a result, industrial systems tend to under perform considering their hardware hardware capabilities. With the growing demand for accelerators for medical sterilization, food irradiation, cancer treatment, and imaging, improving the signal processing of these machines will increase the margin for the deployment of these systems. Our work is focusing on using machine learning techniques to reduce the noise of RF signals used for pulse-to-pulse feedback in industrial accelerators. We will review our algorithms, simulation results, and results working with measured data. We will then discuss next steps for deployment and testing on an industrial system.

\end{abstract}

\section{Introduction}

Machine learning (ML) has been identified as having the potential for significant impact on the modeling, operation, and control of particle accelerators (e.g. see~\cite{7454846, Edelen:2018jid}). In the machine diagnostics space specifically, there have been numerous efforts aimed at improving measurement capabilities and detecting faulty instruments. For example, many developments have focused on the improvement of beam position monitors (BPMs), including the removal of poorly performing BPMs. Work done at the Large Hadron Collider (LHC) identified faulty BPMs prior to application of standard optics correction algorithms~\cite{fol2019jacow}. More recently, ML methods have been utilized to improve optics measurements from beam position monitor data~\cite{fol2018machine}. 

While machine learning continues to be a popular area of research for BPM and other accelerator diagnostics, there is a real dearth of engineering knowledge when it comes to the application of machine learning for RF systems. As the demand for industrial accelerators increases, so does the complexity of these systems and the need for tighter control on the RF system. Machine learning has potential to improve accelerator operations especially for systems operating in an industrial environment. The ability to improve signal to noise ratio and extract key characteristics from RF signals would greatly improve the ability of industrial systems to meet the growing performance demands.  

Autoencoders are a machine learning technique that is well established for the removal of noise from diagnostic signals. Variational Autoencoders (VAEs) are especially adept at removing noise due to the enforcement of a smoothness criterion in the latent-space \cite{higgins2017betavae}. This feature of VAEs has been applied to gravitational wave research~\cite{PhysRevResearch.2.033066,PhysRevD.101.042003} and geophysical data~\cite{bhowick2019stacked}. Recurrent autoencoders have the added advantage of being well suited to work with data sequences. Autoencoders have also been applied to BPM data for the automation of noise removal~\cite{edelen:ibic2022-mop10}. This study examined not just white noise but also the removal of different power law spectra (colors) of noise from simulated BPM data in a ring.

In this paper we evaluate various machine learning methods for noise removal and compare them with a more conventional approach using Kalman filters. Here we begin with a review of our data generation model followed by an analysis of Kalman filters, feed-forward autoencoders, convolutional autoencoders, and variational autoencoders for the removal of noise from RF signals. We then compare the results of these methods and evaluate convolutional neural networks on measurement data collected on an industrial RF system.

\section{Data Generation}
Our data was generated using an RF simulator that reproduces waveforms as they would be seen in industrial systems. Over the past year, RadiaSoft has been developing a full RF simulation tool that is integrated with EPICS for the development of new control algorithms, developing IOC software, and testing user interfaces. The simulator can be run through various APIs including a command line interface, via a Jupyter notebook, or directly through an EPICS connection. The simulator is based on a linear circuit model that takes into account coupling factor, quality factor, frequency, drive amplitude and phase, pulse duration, detuning, etc. The dynamics of our model are based off of equations derived in~\cite{CZARSKI2006565, 10.1063/1.5041079, Echevarria:2018gjw}.

The data were generated by varying the RF pulse characteristics and the cavity characteristics. The RF frequency of the cavity for this study was 2856 MHz, a typical frequency for industrial applications. The pulse length was varied from 3 $\mu$s to 7.5 $\mu$s which is a reasonable range for industrial accelerator applications operating at S-Band.  Additionally we varied the start time of the RF pulse in the data window. While we typically don't expect the RF pulse to vary in position along the DAQ window, adding in this flexibility will ensure better generalization when transferring from simulations to measurement. 

The RF cavity parameters of interest for this study are $Q_0$ and $\beta$ which were varied over a range of 10,000 to 225,000 for the $Q_0$ and 1 to 3 for $\beta$. The detuning was also varied within a range of plus minus one half bandwidth, a fairly typical range seen on industrial systems. In all, the parameter range chosen represents a reasonable range of industrial RF systems and will allow us to develop simulation based algorithms that should be readily transferable to measurement when the time arises.

\section{Kalman Filters}

First we consider the standard Kalman filter \cite{10.1115/1.3662552} for dynamical state estimation. Standard Kalman filters, a class of linear quadratic estimators, predict the state of a system given a prediction for a previous state, a linear model of the dynamics, and all inputs to the system. It then performs a correction step using Bayesian statistics to assimilate incoming data and update the estimator. The model used is comprised of the 1-D dynamical equations for an RF cavity derived from an equivalent circuit model. In  this context, the state, control, and measurement vectors in the linear dynamical model for the cavity dynamics are given in Equation \ref{dyn_var}. 
\begin{equation}
\label{dyn_var}  
\mathbf{x} = \left[ \begin{tabular}{c}
 $\Re(V_t) $\\
 $\Im(V_t) $ \end{tabular} \right] 
\hspace{2mm}
 \mathbf{u} = \left[ \begin{tabular}{c}
 $\Re(I_{fw}) $\\
 $\Im(I_{fw}) $ \end{tabular} \right] 
 \hspace{2mm}
 \mathbf{y} = \left[ \begin{tabular}{c}
 $\Re(V_t) $\\
 $\Im(V_t) $ \\
 $\Re(V_r) $ \\
 $ \Im(V_r)$ \end{tabular} \right] 
\end{equation} 
The dynamical equations that describe the system with noise are given by $\dot{\mathbf{x}} = A \mathbf{x} + B \mathbf{u} + \Gamma \tilde{\mathbf{w}}$ and $\mathbf{y} = C \mathbf{x} + D \mathbf{u} + \tilde{\mathbf{v}}$, where $\tilde{\mathbf{w}}$ and $\tilde{\mathbf{v}}$ are the noise components that show up in the dynamics and that we wish to remove. The matrices $A$, $B$, $C$, and $D$, are defined by the cavity dynamics model as:
\begin{equation} A = \left[\begin{tabular}{c c} $-\omega_{1/2} $ & $\Delta\omega $  \\   $\Delta \omega$ & $  -\omega_{1/2}$  \end{tabular} \right] \end{equation} 
\begin{equation}
B= {R_L \omega_{1/2} \over m} \left[\begin{tabular}{c c} $1 $ & $ 0 $  \\   $0$ & $ 1$  \end{tabular} \right] 
\end{equation} 
 \begin{equation} C= \left[\begin{tabular}{c c} $1 $ & $ 0 $  \\   $0$ & $ 1$ \\ $ 1/m$ & $0 $\\ $0$ & $1/m$  \end{tabular} \right] 
\end{equation} 
\begin{equation}  D= {Z_0 \over 2} \left[\begin{tabular}{c c} $0 $ & $ 0 $  \\   $0$ & $ 0$ \\ $1 $ & $ 0 $  \\   $0$ & $ 1$ \end{tabular} \right] \end{equation}.

\begin{figure*}[!ht]
    \centering
    \includegraphics[width = \textwidth] {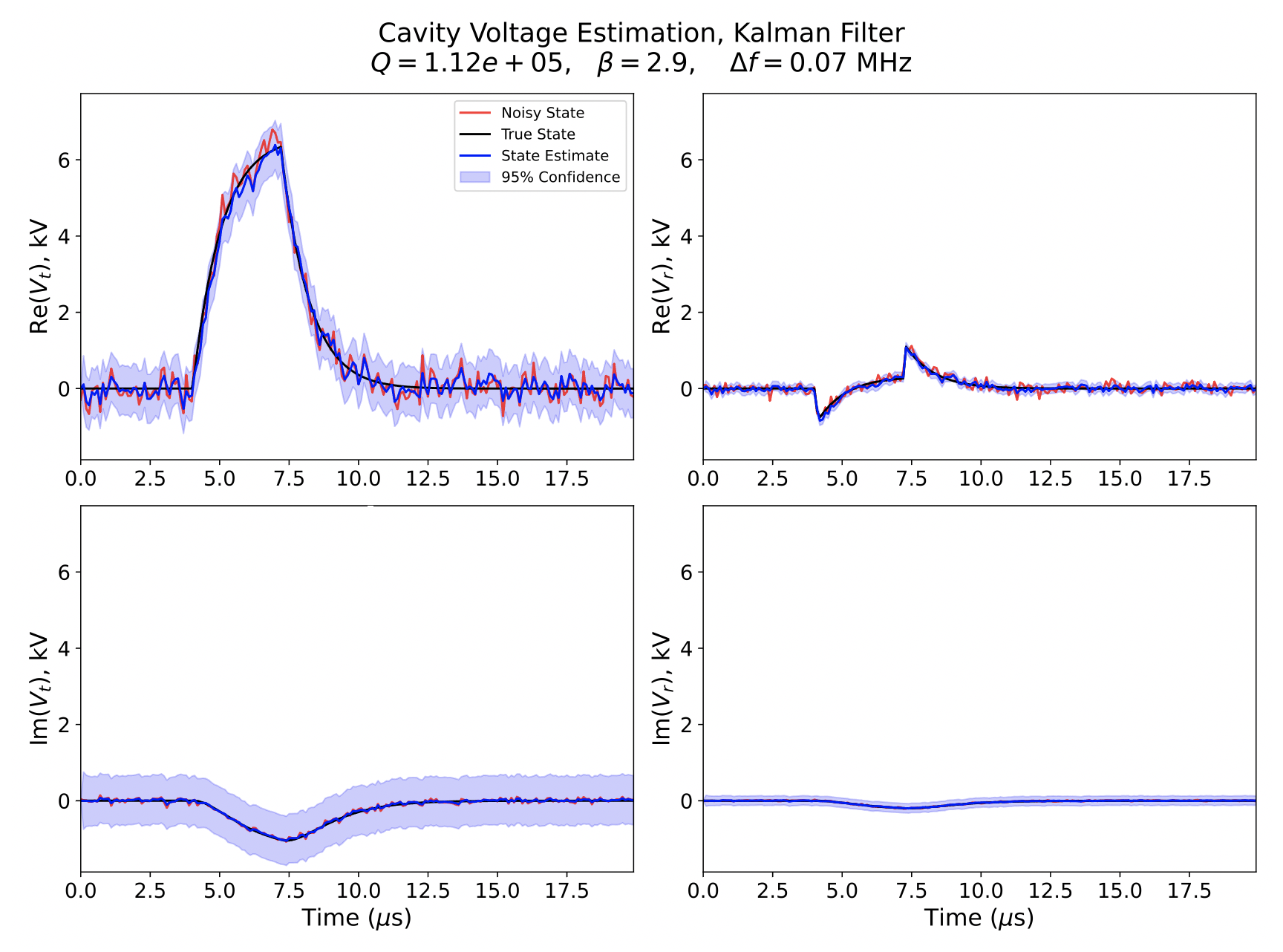}
    \caption{Comparison of the reconstructed waveform from the Kalman filter (red) }
    \label{fig:kal1}
\end{figure*}

This continuous-time representation can be transformed into a discrete representation and then used to define update formulae that allow us to estimate the true system response in the absence of noise. Additionally, the Kalman filter algorithm produces an estimate error covariance matrix which gives us an uncertainty metric on the prediction in addition to the denoised data. Figure \ref{fig:kal1} shows an example waveform result of the Kalman filter. Here we predict the both the transmitted and reflected signals in both I and Q domains. 

By its nature, the Kalman filter used in this way can be seen to have reasonable denoising capabilities. Here the output of the Kalman filter is shown in blue with the confidence interval in shaded blue. The ground truth and noisy signal are in black and red, respectively. Compared with the baseline noisy signal, the Kalman filter does quite well (especially considering it does not require training). One downside to this specific Kalman filter approach is that the forward current constitutes a known input and does not get estimated or denoised. 

\section{Feed-forward Autoencoder}

\begin{figure}[h!]
    \centering
    \includegraphics[width = 0.5\textwidth] {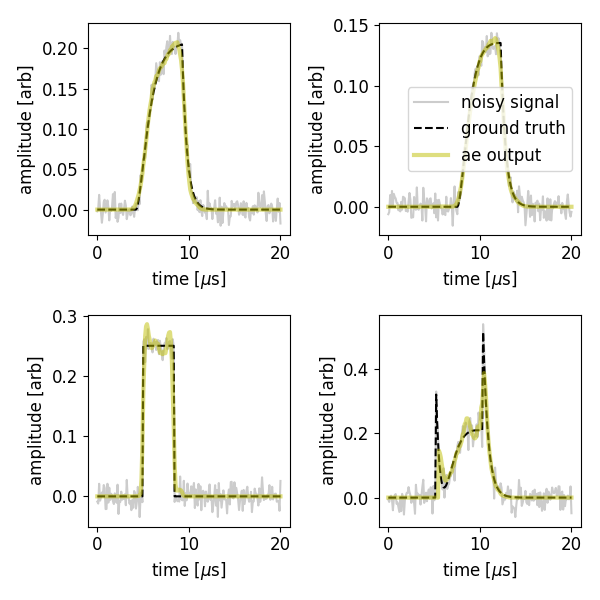}
    \caption{Comparison between the autoencoder reconstruction (yellow) of four randomly chosen waveforms and the ground truth (black) and the noisy signal (gray)}
    \label{fig:ae1}
\end{figure}

Of the neural network techniques, we first utilized feed-forward neural networks for noise reduction. Here the input data are the individual RF signals where each time-step in the signal is a feature in the input space. We trained a different autoencoder on each type of waveform for the forward, reflected, and probe signals. The autoencoder was evaluated on the test dataset which is identical for each of the methods explored in this study. The architecture of our model was relatively simplistic with a single encoder layer and a latent dimension of 32. The model was trained using noisy waveform data from our simulator in an unsupervised fashion. That is during training the model inputs and outputs both contain noise. The mechanism for noise reduction is due to the fact that there is not enough complexity in the latent space to reconstruct the noise. Figure \ref{fig:ae1} shows the result of the feed-forward autoencoder for noise reduction of four example waveforms.

\section{Convolutional Autoencoder}

\begin{figure}[h!]
    \centering
    \includegraphics[width = 0.5\textwidth] {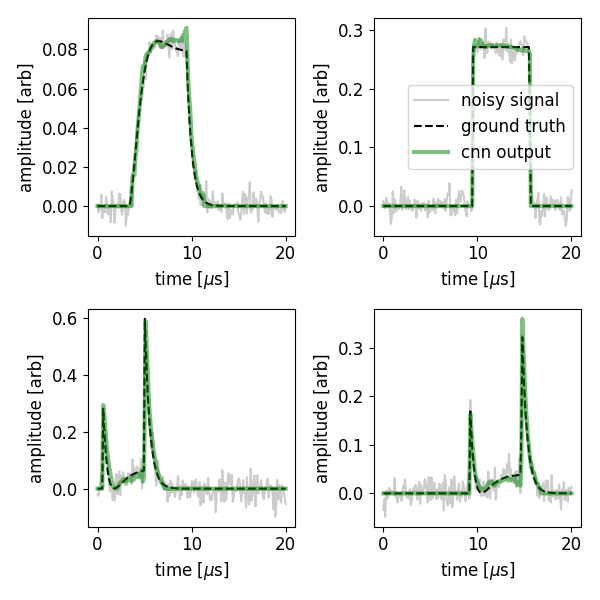}
    \caption{Comparison between the 1D-CNN reconstruction (green) of four randomly chosen waveforms and the ground truth (black) and the noisy signal (gray)}
    \label{fig:cnn1}
\end{figure}

Convolutional neural networks (CNNs) are adept at feature extraction especially in cases where there is translation invariance. While typical LLRF signals are time synchronized, we explored signal translations as described above to improve the generality of our approach. The convolutional network follows a structure very similar to a U-net which is often used for image segmentation and other image to image learning problems. The model architecture consisted of 1-D convolutional layers and max-pooling layers that reduce the feature space down to a latent space of 10. We then used up-sampling and convolutional layers to reconstruct the waveform. When training the CNN, we treated each waveform as unique to allow the CNN to learn noise rejection regardless of if the data being processed is a forward, reflected, or probe signal. This will improve our ability to generalize when considering data collected on different types of machines where probes are not always available or traveling wave structures where the signal envelopes do not follow the same profile as standing wave cavities. Figure~\ref{fig:cnn1} shows the noisy signal in grey, the model prediction in green, and the ground truth signal with no noise in black. 

For comparison, we examine drive, reflected, and cavity probe signals. While the reconstructed signal is generally closer to the ground truth than the noisy signal, there are cases where spurious signals are present in the reconstructed data (top left).

\section{Variational Recurrent Autoencoders}

Next we considered variational recurrent autoencoders, VRAEs. VRAEs are an excellent tool for data reduction and noise elimination due to the fact that they enforce a smoothness criterion in the latent space while also utilizing recurrent layers to effectively translate time dynamics to the principal components of the simulation which will be represented in the latent space.  

Our implementation of the VRAE architecture is based on~\cite{fabius2015variational} and uses Long Short-Term Memory (LSTM) units for both the encoder and decoder. The loss function is composed of two terms: the Kullback–Leibler divergence~\cite{kingma2014autoencoding} --- which acts as a regularization term --- and the reconstruction loss. For the reconstruction loss, mean squared error between the encoder input and decoder output is used. The VRAE is trained and tested with each of the waveforms (drive, reflected, and probe) treated as features for the dataset. The goal here is that the VRAE will be able to learn a latent space representation of the waveform data and by extension the cavity model parameters. This is in contrast to the CNN which was trained to process one waveform at a time and is largely learning to remove noise as opposed to a more generalized representation of the system. 

\begin{figure}[h!]
    \centering
    \includegraphics[width = 0.5\textwidth] {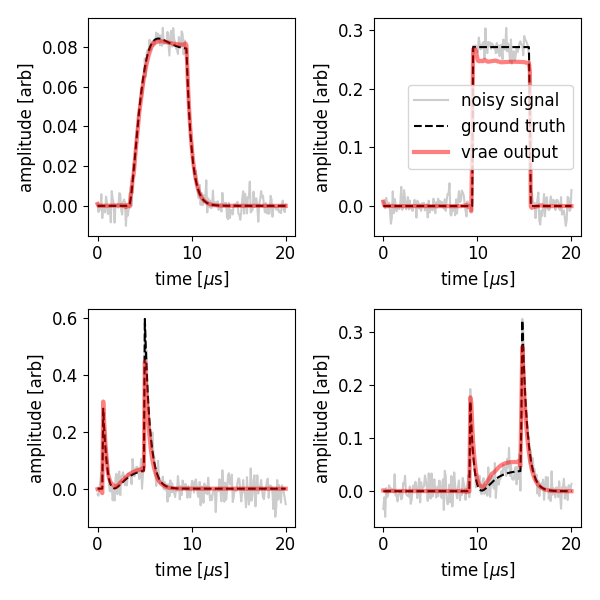}
    \caption{Comparison between the VRAE reconstruction (red) of four randomly chosen waveforms and the ground truth (black) and the noisy signal (gray)}
    \label{fig:enter-label}
\end{figure}

The prediction for the VRAE is generally quite a bit better in terms of the ability to remove noise, but the profile of the reconstructed waveform is not always correct. This is likely due to the fact that there is some degeneracy in the dataset due to the relationship between $Q$ and $\beta$. 

\section{Comparison}

\begin{figure}[h!]
    \centering
    \includegraphics[width = 0.5\textwidth] {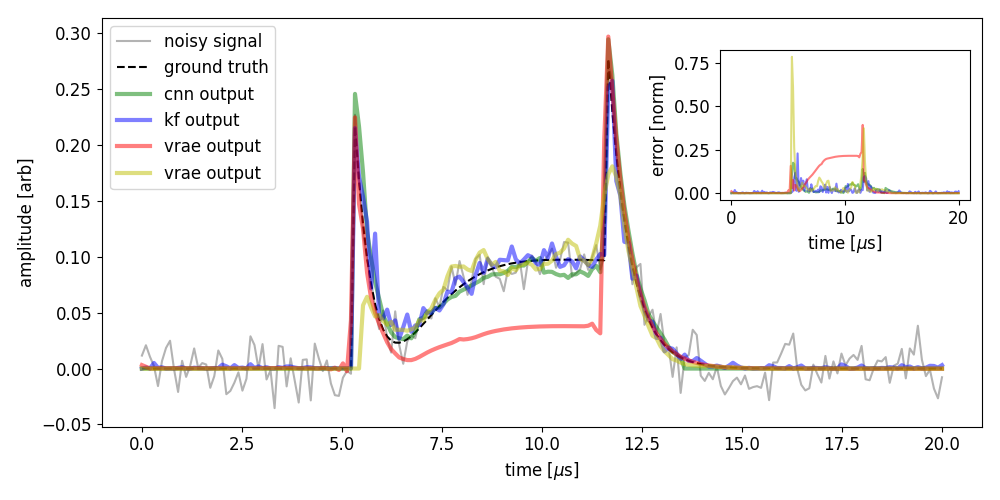}
    \caption{Reflected power waveform reconstruction using the Kalman Filter (blue), the 1D-CNN (green) and the VRAE (red) with the respective errors shown in the inset plot.}
    \label{fig:comp1}
\end{figure}

\begin{figure}[h!]
    \centering
    \includegraphics[width = 0.5\textwidth] {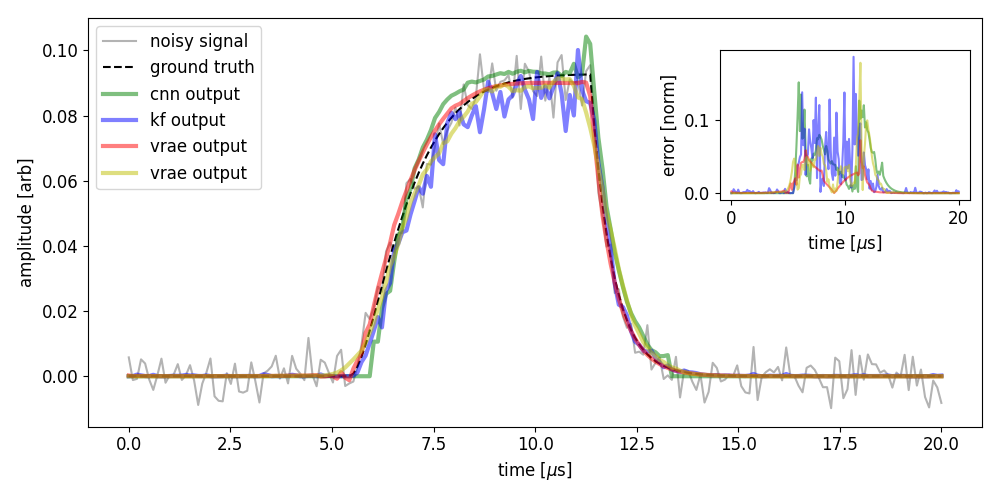}
    \caption{Cavity power waveform reconstruction using the Kalman Filter (blue), the 1D-CNN (green) and the VRAE (red) with the respective errors shown in the inset plot.}
    \label{fig:comp2}
\end{figure}

In all four cases the approaches are capable of reducing noise in the waveform data. Figures \ref{fig:comp1} and \ref{fig:comp2} show a direct comparison between these approaches and the ground truth in addition to the original noisy signal. Here we can see that the CNN consistently does well compared to the Kalman filter. The VRAE does quite well on the second example but does not reconstruct the signal properly for the first example. The Kalman filter has several advantages, in that it both performs some noise reduction while also relying on the physics of the system without training. This leads to the Kalman filter being less likely to produce spurious signals, such as those seen in the second example using the CNN right before the cavity turns off. 

\begin{figure}[h!]
    \centering
    \includegraphics[width = 0.5\textwidth] {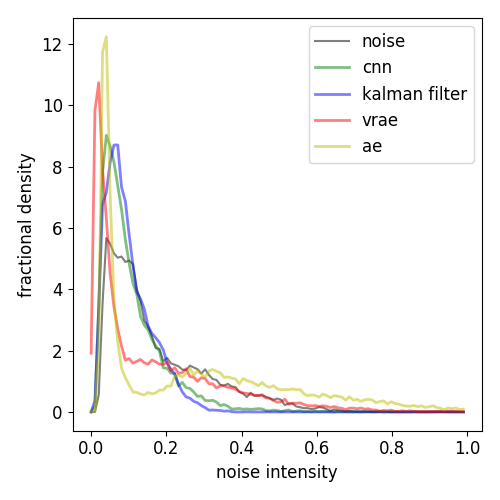}
    \caption{Histogram of the sum-squared-error between the reconstructed signal and the noiseless signal for the autoencoder (yellow), VRAE (red), 1D-CNN (green), and Kalman filter (blue).}
    \label{fig:comp3}
\end{figure}

To evaluate the performance of each technique across the whole dataset, we computed the sum squared error between the reconstructed signal and the ground truth for each example waveform. We then computed a histogram of these results and compared it to histogram of the sum-squared-error between noisy signal and the noiseless signal. Figure~\ref{fig:comp3} shows the result of this comparison. 

The comparison across the whole test dataset shows that each method is capable of reducing the noise levels of the signals. The VRAE does the best job as can be seen by the prominent spike in the histogram near zero and a significant reduction in the noise contribution between 0.1 and 0.2. The CNN and the Kalman filter perform similarly overall with the CNN showing slightly better noise reduction. 

\begin{figure}[h!]
    \centering
    \includegraphics[width = 0.5\textwidth] {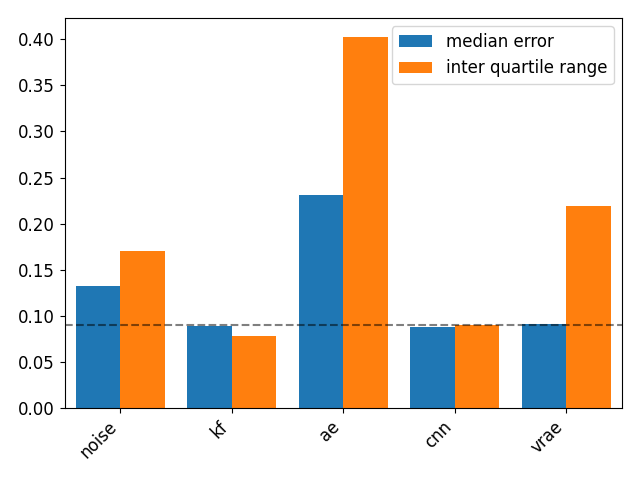}
    \caption{Median and interquartile range for the root-sum-squared-error of the test data.}
    \label{fig:comp4}
\end{figure}

Figure~\ref{fig:comp4} shows a comparison of the root-sum-squared-error for each of the methods. Here we compute the median and interquartile range for each method. The Kalman Filter and CNN outperform the VRAE in the interquartile range despite fact that the medians for the VAE, AE, and CNN are virtually indistinguishable. This shows that they perform well across a broader subset of the data compared to the VAE. 

\begin{figure}[h!]
    \centering
    \includegraphics[width = 0.5\textwidth] {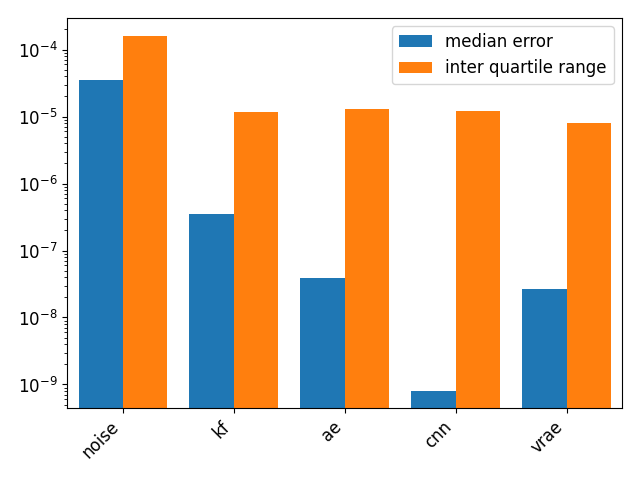}
    \caption{}
    \label{fig:comp5}
\end{figure}

We also compared the median and interquartile range of the squared error across the whole dataset to get a broader picture of how the different methods compare. Figure \ref{fig:comp5} shows the median and interquartile range of the squared error on the test set. This shows that each method removes a significant amount of noise as evidenced by the orders of magnitude smaller median errors for all four techniques. The interquartile range is also significantly reduced but here each method is more comparable. This indicates that although nose is being removed, the spread in the error is still relatively large. 

\section{Retraining on Measurement Data}

\begin{figure*}[h!]
    \centering
    \includegraphics[width =\textwidth] {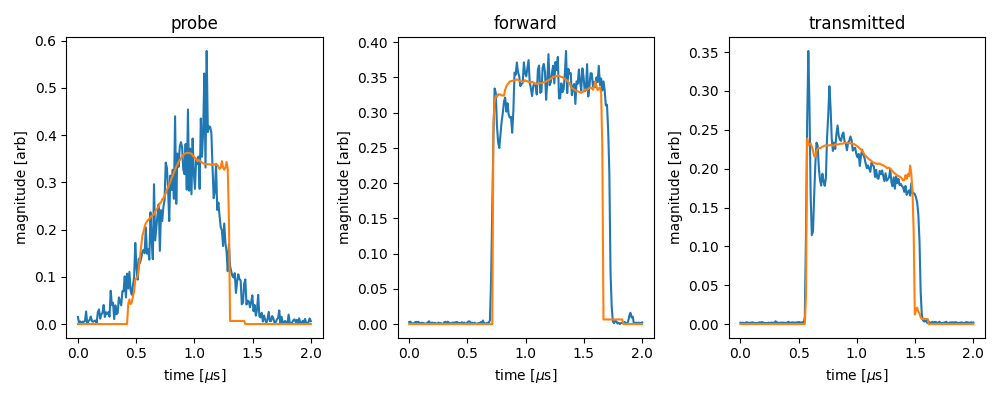}
    \caption{Reconstruction and input signal for three measurement waveforms with significant noise taken from RadiaBeam.}
    \label{fig:meas1}
\end{figure*}

Finally we tested the CNN autoencoder on measured data from a cavity being tested at RadiaBeam. This structure is novel and does not have the same RF characteristics as the training data. Making it an ideal test for our ML methods. Because the sample rate of the test data is much larger than our simulation data we down-sampled the data prior to feeding it to the neural network. The data were also scaled to be in a range of 0-1 which is the same range of our training data. Figure\ref{fig:meas1} shows the input waveform (blue) and the reconstruction (orange) for three RF waveforms collected from a test cavity. 

\begin{figure*}[h!]
    \centering
    \includegraphics[width =\textwidth] {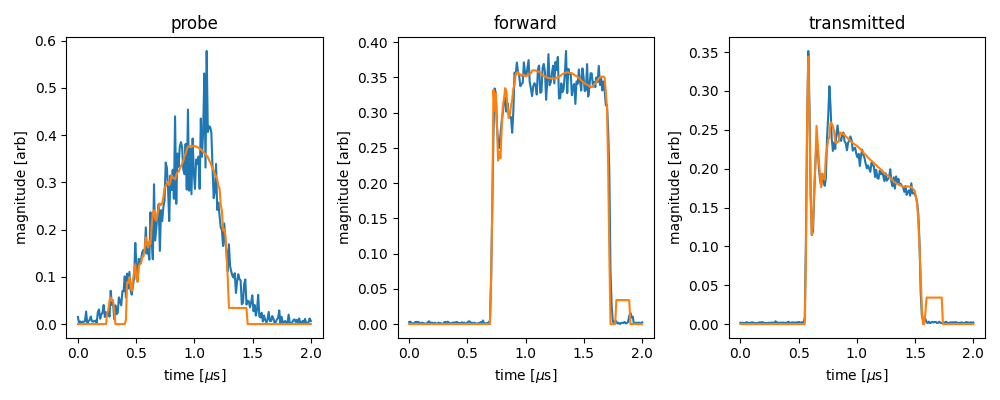}
    \caption{Reconstruction and input signal for three measurement waveforms with significant noise taken from RadiaBeam after retraining.}
    \label{fig:meas2}
\end{figure*}

Here we can see that while the overall shape of the reconstruction is correct certain key aspects are missing. The ringing in both the forward and transmitted signals are not reconstructed while there appears to be a ringing in the forward signal that is reconstructed where there is really just noise. Moreover the probe signal is not well reconstructed at all. Note that the pulse length and cavity frequency are outside the domain of our training data. We performed a batch retraining using a small subset of the data (24 waveforms) and then examined the performance of the CNN after retraining. Figure \ref{fig:meas2} shows the resulting waveform reconstructions after retaining. 

Here the reconstruction does capture the initial ringing in both the forward and the transmitted signals while not introducing significant oscillations in the transmitted signal during the down-ramp. In fact the transmitted signal reconstruction is near perfect from a noise elimination standpoint. The probe signal however still has a large degree of variation as opposed to a clean envelope. 

\section{Conclusions}

We have explored four possible approaches for noise reduction on industrial RF signals using machine learning. Specifically we have utilized Kalman filters, autoencoders, convolutional autoencoders, and variational autoencoders. Our test dataset was generated using a cavity circuit model that has been well benchmarked against measured data. While each method has significant noise reduction capabilities, the VRAE and the CNNs are generally the best tool for this task. We tested our method on measurements from a test cavity at RadiaBeam using the CNN. The prediction before retraining leaves much to be desired while after only a amount of retraining the predictions are much closer to an ideal reconstruction of the waveform. 

\section{Acknowledgements}

This material is based upon work supported by the U.S. Department of Energy, Office of Science, Office of Accelerator R\&D and Production Award Number DE-SC0023641.

\ifboolexpr{bool{jacowbiblatex}}%
    {\printbibliography}%
    {%
    % "biblatex" is not used, go the "manual" way
    
    %\begin{thebibliography}{99}   % Use for  10-99  references
    
} % end \ifboolexpr

\end{document}